\documentclass[12pt,preprint]{aastex}
\makeatletter

\newcommand{\Rmnum}[1]{\expandafter\@slowromancap\romannumeral#1@}
\makeatother
\usepackage{amsmath,amssymb}
\slugcomment{To be submitted to ApJL} 
\shorttitle{2D simulations of prominence formation}
\shortauthors{Xia et al.}
\begin{document}
\title{SIMULATIONS OF PROMINENCE FORMATION IN THE MAGNETIZED SOLAR 
       CORONA BY CHROMOSPHERIC HEATING}
\author{C. Xia\altaffilmark{1}, P. F. Chen\altaffilmark{1}, 
R. Keppens\altaffilmark{2}}
\altaffiltext{1}{School of Astronomy and Space Science, Nanjing University, 
Nanjing 210093, China; chenpf@nju.edu.cn}
\altaffiltext{2}{Centre for Plasma Astrophysics, Department of 
Mathematics, K.U.Leuven, Celestijnenlaan 200B, 3001 Heverlee, Belgium}

\begin{abstract}
Starting from a realistically sheared magnetic arcade connecting 
chromospheric, transition region to coronal plasma, we simulate the 
in-situ formation and sustained growth of a quiescent prominence in 
the solar corona. Contrary to previous works, our model captures all 
phases of the prominence formation, including the loss of thermal 
equilibrium, its successive growth in height and width to macroscopic 
dimensions, and the gradual bending of the arched loops into dipped 
loops, as a result of the mass accumulation. Our 2.5-dimensional, 
fully thermodynamically and magnetohydrodynamically consistent model 
mimics the magnetic topology of normal-polarity prominences above a 
photospheric neutral line, and results in a curtain-like prominence 
above the neutral line through which the ultimately dipped magnetic 
field lines protrude at a finite angle. The formation results from 
concentrated heating in the chromosphere, followed by plasma 
evaporation and later rapid condensation in the corona due to thermal 
instability, as verified by linear instability criteria. Concentrated 
heating in the lower atmosphere evaporates plasma from below to 
accumulate at the top of coronal loops and supply mass to the later 
prominence constantly. This is the first evaporation-condensation 
model study where we can demonstrate how the formed prominence stays 
in a force balanced state, which can be compared to the 
Kippenhahn--Schl\"{u}ter type magnetohydrostatic model, all in a 
finite low-beta corona.
\end{abstract}


\section{INTRODUCTION}\label{intro}
Prominences represent fascinating large-scale, cool ($\sim$ 8000 K) 
and dense ($10^{10}\sim10^{11}$ cm${^{-3}}$) structures, suspended 
in the hot and tenuous solar corona above magnetic neutral lines, 
which separate opposite polarity photospheric magnetic regions. The 
magnetic field strength in quiescent prominences lies between 3--15 G 
\citep{Lero83}. It is much stronger in active region prominences. The
field topology of prominences is less accurately known, but was found 
to be mostly horizontal, making an acute angle with respect to the 
axis of the prominence \citep{Bomm98}. It is accepted that the stable 
existence of prominences is due to the mechanical support and thermal 
shielding of the magnetic fields. Two topologically different types 
of static prominence models were proposed by \citet{Kipp57} and 
\citet{Kupe74}. In the Kippenhahn--Schl\"{u}ter (K--S) prominences, 
also known as normal-polarity prominences, the horizontal magnetic 
field through the prominence points from the underlying positive to 
the negative magnetic polarities in the photosphere. In Kuperus--Raadu 
(K--R) prominences, or inverse-polarity prominences, the magnetic 
field points from negative to positive. In both types, the 
concave-upward parts of magnetic field lines or `dips' host and 
support prominence material via the magnetic tension force against 
gravity. These have been extended by many authors \citep{Low75,Amar89,
Hood90}, and recently, flux-rope embedded, normal- and inverse- 
polarity type equilibrium configurations have been amenable to 
numerical modeling \citep{Petr07,Blok11}. These models merely 
consider magnetohydrostatic force balance arguments, and leave out 
the thermodynamic aspects.

The most elusive aspect in prominence physics is their seemingly 
in-situ formation. One of the earliest suggested formation scenarios
relates the sudden appearance of prominence material to a radiative 
condensation process due to thermal instability \citep{Park53,Fiel65}. 
In the optically thin solar corona, the radiative losses are 
proportional to density squared and the temperature can evolve into 
the critical range where a decrease of temperature leads to an 
increase of the radiative losses \citep{Colg08}. When the plasma is 
perturbed to become dense or cool enough for radiative losses to 
dominate both the heating by thermal conduction and any background 
heating process, sudden strong cooling can cause sufficient pressure 
gradient to suck in material around along field lines. Hence, plasma 
condenses into cool regions and settles in magnetic dips to form 
large-scale prominences. The linear `trigger' of such 
thermal-instability condensations was studied in multi-dimensional 
sheared magnetic field by many authors \citep{vanH84,vand92}. The 
nonlinear evolution of such thermal instability can only be 
researched by numerical simulations. \citet{Hild74} firstly 
performed two-dimensional (2D) numerical simulations of the thermal 
instability for the formation of prominences in a uniform magnetic 
and gravitational field, while neglecting thermal conduction. 
\citet{Spar90} simulated a nonlinear condensation in a force-free 
sheared magnetic field including anisotropic thermal conduction, but 
neglected gravity. \citet{Drak93} simulated the formation and stable 
support of a prominence at the apex of a potential magnetic arcade 
including the upper chromosphere. But in their simulations, the 
initial trigger perturbations were artificially added without any 
detailed mechanism ultimately giving rise to condensation into a 
prominence.

Many mechanisms for the transport of plasma from the chromosphere to 
the corona thereby triggering the formation of prominences have 
meanwhile been proposed\citep[see the review by][]{Mack10}. 
\citet{An88} and \citet{Wu90} simulated the formation of a prominence 
in a potential field arcade in an initially isothermal corona by 
injecting high density material ballistically from the bottom 
boundary. \citet{Choe92} performed a 2.5-dimensional (2.5D) 
magnetohydrodynamic (MHD) simulation to investigate prominence 
formation due to siphon flows by photospheric shearing motions. These
simulations did not include photosphere or chromosphere regions. 
One-dimensional (1D) evaporation-condensation model, which include
chromosphere and chromospheric heating, have been studied numerically 
by many authors \citep{Mok90,Anti99,Karp01,Xia11}. This 
thermo-hydrodynamic model depends on heating localized in the 
chromospheric feet of a pre-shaped loop, which evaporates 
chromospheric material into the corona. The increase in density 
results in a dominating radiative cooling, and the thermal 
instability triggers condensations in the corona, eventually forming 
a steadily growing or recycling prominence \citep{Xia11}. In this 
Letter, we present the first simulation where this model is performed 
in a 2.5D full MHD regime in a realistically stratified sheared 
arcade going from the low chromosphere to high corona. We show how 
we can simulate the rapid thermodynamical changes in a scenario where 
we form and reach a stable support for a K--S quiescent prominence.

\section{NUMERICAL SETUP}\label{numer}
We perform our simulation in a rectangular, Cartesian geometry, 
ignoring the curvature of the solar surface. Since prominences are 
often observed in a horizontally elongated form along the magnetic 
neutral line, we ignore the variation of physical variables along the 
prominence axis ($z$-direction) and consider their dependence on 
horizontal $x$-coordinate (perpendicular to the magnetic neutral line) 
and vertical $y$-coordinate, while keeping the $z$-component of 
any vector. Therefore, we use a 2.5D thermodynamic MHD model 
including gravity, field-aligned heat conduction, radiative cooling 
and heating terms. The governing equations are as follows
\begin{equation}
 \frac{\partial \rho}{\partial t}+\nabla\cdot\left(\rho\mathbf{v}
   \right)=0,
\end{equation}
\begin{equation}
 \frac{\partial \left(\rho\mathbf{v}\right)}{\partial t}+\nabla\cdot
  \left(\rho\mathbf{vv}+p_{tot}\mathbf{I}-\mathbf{BB}\right)=\rho
  \mathbf{g},
\end{equation}
\begin{equation}
 \frac{\partial E}{\partial t}+\nabla\cdot\left(E\mathbf{v}+p_{tot}
  \mathbf{v}-\mathbf{BB}\cdot\mathbf{v}\right)=\rho\mathbf{g}\cdot
  \mathbf{v}+\nabla\cdot\left(
  \boldsymbol{\kappa}\cdot\nabla T\right)-Q+H,
\end{equation}
\begin{equation}
 \frac{\partial \mathbf{B}}{\partial t}+\nabla\cdot\left(\mathbf{vB}
  -\mathbf{Bv}\right)=0,
\end{equation}
where $\rho$, $T$, $\mathbf{v}$, $\mathbf{B}$, and $\mathbf
{I}$ are the plasma density, temperature, velocity, magnetic field, 
and unit tensor, respectively; $p_{tot}\equiv p+B^2/2$ is the total 
pressure, composed of thermal pressure $p$ and magnetic pressure
$B^2/2$; $E=p/(\gamma-1)+\rho v^2/2+B^2/2$ is the total energy 
density, where $\gamma=5/3$ is the ratio of specific heats; 
$\mathbf{g}=-g_0R_\odot^2/(R_\odot+y)^2\mathbf{\hat{y}}$ is the 
gravitational acceleration with $R_\odot$ the solar radius and $g_0$ 
the solar surface gravitational acceleration; $\boldsymbol{\kappa}$ 
is the thermal conductivity tensor; $Q$ and $H$ are the radiative loss 
rate and heating densities, respectively. Considering fully ionized 
plasma with 10:1 abundance of hydrogen and helium, we have $\rho=1.4 
m_p n_{\rm H}$, where $m_p$ is the proton mass and $n_{\rm H}$ is the 
number density of hydrogen. As for the equation of state, we adopt the 
ideal gas law $p=2.3 n_{\rm H}k_B T$, where $k_B$ is the Boltzmann 
constant. The radiative cooling term is taken as $Q=1.2 n_{\rm H}^2
\Lambda(T)$, where $\Lambda(T)$ is the radiative loss function for 
optically thin emission, which was also used in our earlier 1D studies 
\citep{Xia11}. The anisotropic thermal conduction along the magnetic 
field lines is included with 
$\boldsymbol{\kappa}=\kappa\mathbf{\hat{b}\hat{b}}$, where 
$\mathbf{\hat{b}}$ is the unit vector along $\mathbf{B}$ and 
$\kappa=10^{-6}T^{5/2}$ erg cm$^{-1}$ s$^{-1}$ K$^{-1}$ is 
the Spitzer conductivity.

As for the initial magnetic field, we intend to mimic sheared arcades 
above a neutral line ($x=0$, $y=0$) and start our simulation from a 
force-balanced state. Therefore we adopt an analytic solution of a 
nonlinear force-free field found by \citet{Low77} as follows
\begin{equation}
 B_x=2 B_0 \left(ky+\frac{1-\mu^2}{1+\mu^2}\right)/f,~~~~ B_y=-2 B_0 k
 x/f, ~~~~B_z=\frac{4\mu B_0}{1+\mu^2}/f,
\end{equation}
\begin{equation}
 \textrm{ where~}f=\frac{4\mu^2}{(1+\mu^2)^2}+k^2x^2+
  \left(ky+\frac{1-\mu^2}{1+\mu^2}\right)^2,
\end{equation}
where $\left|\mu\right|<1$ controls the shearing rate of the arcades (no
shearing if $\mu=0$) and $k$ controls the spatial concentration of 
the field. The shearing decays from lower loops to higher loops.
We set $\mu=0.95$, $k=0.5$, and $B_0=4$ G, leading to a realistic 
2.5D arcade topology, where the field lines at the height of 20 Mm 
make an angle of 45$^\circ$ with the underlying neutral line. 

For the initial thermal structure, we set a chromosphere with 
temperature of 10000 K below a height of 2.7 Mm and choose a vertically 
stratified temperature profile with a constant vertical thermal 
conduction flux (i.e. $\kappa \partial T/\partial y=2\times10^5$ ergs 
cm$^{-2}$ s$^{-1}$) above the height \citep{Mok05, Font91}. The initial 
density is then determined by assuming a hydrostatic atmosphere with 
the number density of $2\times10^{13}$ cm$^{-3}$ at the bottom. To use 
a proper background heating term to maintain a hot corona, we are 
inspired by parametric comparisons of different models of coronal 
heating done by \citet{Mand00} and assume the heating rate to be
proportional to $B^2$ \citep{Mok08}. However, in order to compensate the
radiative loss in the transition region above the neutral line, where
dominating horizontal field lines insulate this region from getting heat
thermally conducted from the corona above, we add an extra heating 
equal to the local radiative loss, below a 6 Mm height, purely 
concentrated in this region. The resulting two-component, parametric 
background heating is expressed as
\begin{equation}
 H_0=c_0 B^2+0.5 Q \cos\left(\frac{\pi B_y}{2 B_0}\right)
\left[1-\tanh\left(\frac{y-y_h}{y_d}\right)\right],
\end{equation}
with $c_0=7.5\times10^{-6}$ erg cm$^{-3}$ s$^{-1}$ G$^{-2}$, 
$y_h=6$ Mm, and $y_d=0.5$ Mm.

The configuration of this system is symmetric about the $y$-axis. We 
exploit the symmetry to study prominence formation under symmetric 
heating conditions, allowing to reduce computational domain to the 
right half of the simulated area (within $0<x<30$ Mm and $0<y<40$ Mm). 
Symmetric/asymmetric boundary treatments can then be used at the 
$y$-axis, while for right-hand side boundaries we adopt a zero 
velocity, continuous density and pressure, and fixed magnetic field. 
The top and the bottom boundaries have a zero velocity, fixed magnetic 
field and extrapolated density and pressure, respectively, which are 
derived by assuming a hydrostatic equilibrium with the temperature 
being fixed.

We use the parallelized Adaptive Mesh Refinement Versatile Advection
Code (MPI-AMRVAC) \citep{Kepp12} to solve the governing equations with a
second-order shock-capturing Total Variation Diminishing Lax-Friedrichs
scheme. The effective resolution of $512\times1024$ is attained by using
5 levels of AMR. The equivalent spatial resolution is then 39 km/59 km 
in the vertical/horizontal direction. Anisotropic thermal conduction is 
added explicitly as an energy source term.

This initial state is not in thermal equilibrium, and we integrate the
governing equations in time with the background heating $H=H_0$ until 
the system relaxes to a quasi-equilibrium shown by panels (a) and (b)
in Figure~\ref{fig:evo}. The chromosphere of about 4 Mm thickness is 
connected to the corona by a very thin transition region. The projected
magnetic field lines, colored according to the local density, are
plotted through selected footpoints at the bottom and side boundaries.
The plasma beta is 0.1 at 20 Mm height above the neutral line while 
the temperature and number density are 1.6 MK and $2.6\times10^8$ 
cm$^{-3}$ there, respectively. The maximal residual velocity is small, 
less than 5 km s$^{-1}$. Starting from this quasi-equilibrium, a 
relatively strong heating $H_1$ is added. This extra heating is 
localized near the chromosphere (see the contours in Figure 
\ref{fig:evo}(a)) with its formula as:
\begin{equation}
H_1=\begin{cases}C_1 (B_y/B_0)^2& \text{if $y \leqslant y_c$}, \\
 C_1 (B_y/B_0)^2 \exp(-(y-y_c)^2/\lambda)& \text{if $y > y_c$},
 \end{cases}
\end{equation}
where $C_1=10^{-2}$ergs cm$^{-3}$ s$^{-1}$, $y_c=3$ Mm, and 
$\lambda=3$ Mm. This localized heating is concentrated in the regions
of strong $B_y$.

\section{PROMINENCE FORMATION DUE TO EVAPORATION AND THERMAL 
INSTABILITY}\label{results}

\begin{figure}
\includegraphics[width=5.8in]{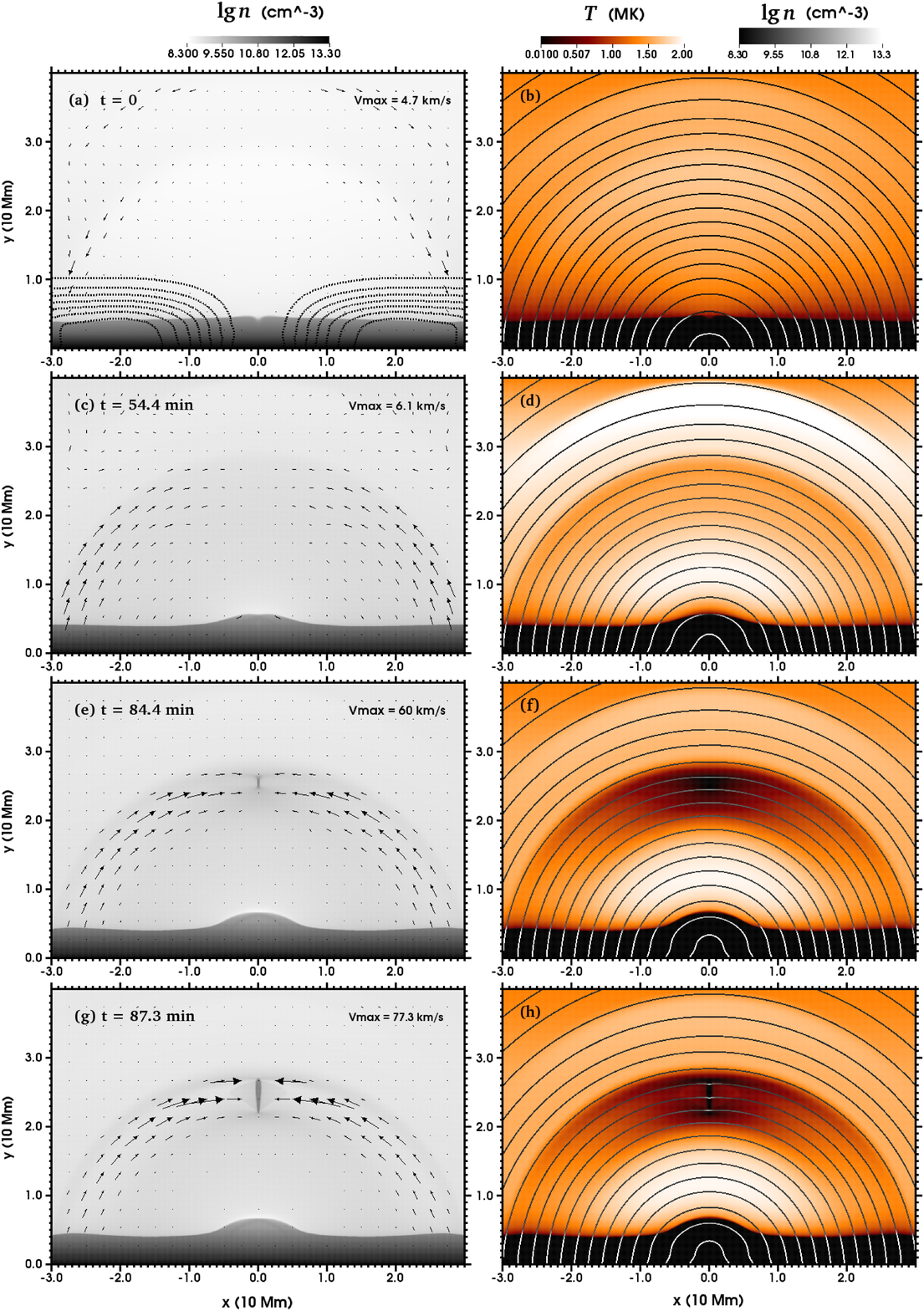}
\caption{Snapshots of the formation process at $t=$ 0 ({\it first row}), 
54.4 mins ({\it second row}), 84.4 mins({\it third row}), and 87.3 mins
({\it bottom row}). In the left column, the density is shown in 
gray and the projected velocity field is shown by arrows. The 
right column shows the temperature and the projected field lines are
colored by the local density. Contours of the localized heating 
are plotted in panel (a) in dotted lines.
}
\label{fig:evo}
\end{figure}

As the localized heating is functioning, chromospheric plasma is heated
and evaporated into the arched coronal loops, increasing the density and
the temperature there. About 6 minutes later, the temperature reaches a 
maximum value of 2.2 MK, and then starts to decrease slowly (see Figure 
\ref{fig:evo}(c, d)). At about 84 minutes, the temperature of an 
inverted triangle-shaped region around the apexes of a bundle of
magnetic loops decreases drastically (see Figure \ref{fig:evo}(f)) and a
small condensation with the typical chromospheric density 
($6.3\times10^{10}$ cm$^{-3}$) appears near the loop top at a height of
25.4 Mm. This is accompanied by two strong inflows moving towards the 
central condensation segment, with a maximum velocity of 70 km s$^{-1}$
from the two sides (see Figure \ref{fig:evo}(e)). The magnetic field
lines near the condensation make an angle of $62^\circ$ with respect to 
the $z$-direction. After this sudden birth of the prominence, 
condensation successively happens on the tops of lower and higher 
coronal loops, leading to a rapid extension of the prominence in the 
vertical direction. As the inflows collide near the loop tops, two 
rebound shock waves are formed (see diamond-shaped wave fronts in 
Figure \ref{fig:evo}(g, h)) and propagate from the apex towards the 
loop feet, during which they sweep across and slow down the evaporated 
upflows. 

\begin{figure}
\includegraphics[width=6.5in]{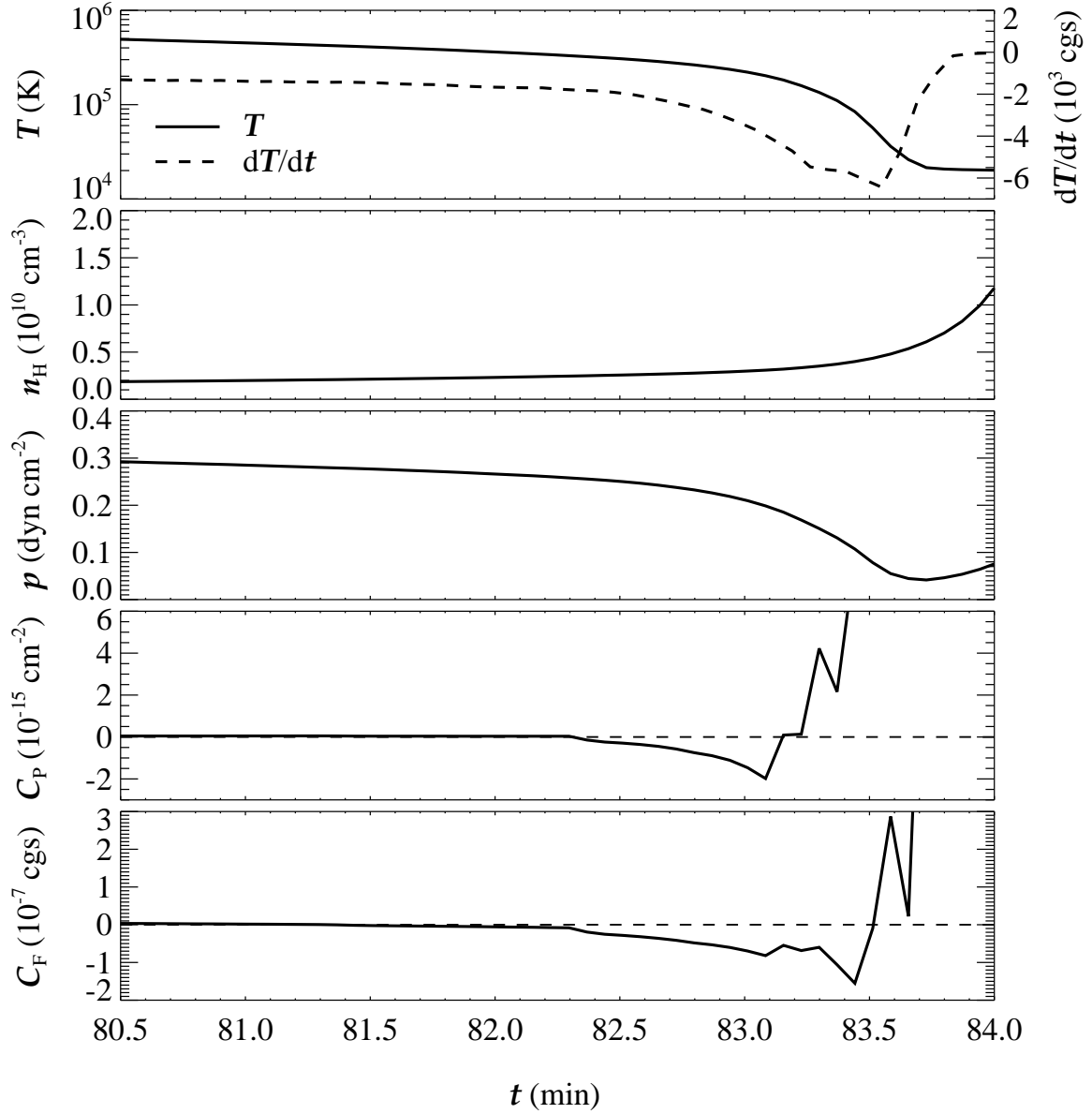}
\caption{Temporal evolutions of the temperature, the density, the
pressure, the thermal instability isochoric criterion $C_P$, and the
isobaric criterion $C_F$ at the first condensation site 
({\it solid lines}), as well as the time derivative of the temperature
({\it dashed line, top panel}). The horizontal dashed lines in the 
last two panels denote the zero value.
}
\label{fig:crit}
\end{figure}

In order to investigate the thermal instability during the in-situ
formation of the prominence, we quantify the temperature, the time 
derivative of temperature, the density, the pressure, the thermal
instability isochoric criterion $C_P$ \citep{Park53} and isobaric 
criterion $C_F$ \citep{Fiel65} at the site ($x$=0, $y$=25.4 Mm) of
the first condensation (see Figure \ref{fig:crit}). A value of 20 Mm 
is adopted as the wavelength of perturbation when calculating the 
thermal instability criteria \citep{Xia11}. After $t$=82.3 minutes, 
when the temperature and pressure start to decrease nonlinearly, both 
$C_P$ and $C_F$ dive into the negative region indicating the 
functioning of thermal instability. Although the localized heating in 
the chromosphere is in a wide range in the $x$-direction, plasma 
condensation appears only on long field loops. The plasma conditions 
along these field loops satisfy the criterion of thermal instability.
At a given height, a larger shear rate (quantified by $\mu$) leads to 
a longer field line whose footpoints are closer to the neutral line, 
which renders the sheared arcade more vulnerable to thermal 
instability.

\section{OVERALL PROMINENCE STRUCTURE AND FORCE BALANCE}\label{discus}
\begin{figure}
\includegraphics[width=6.5in]{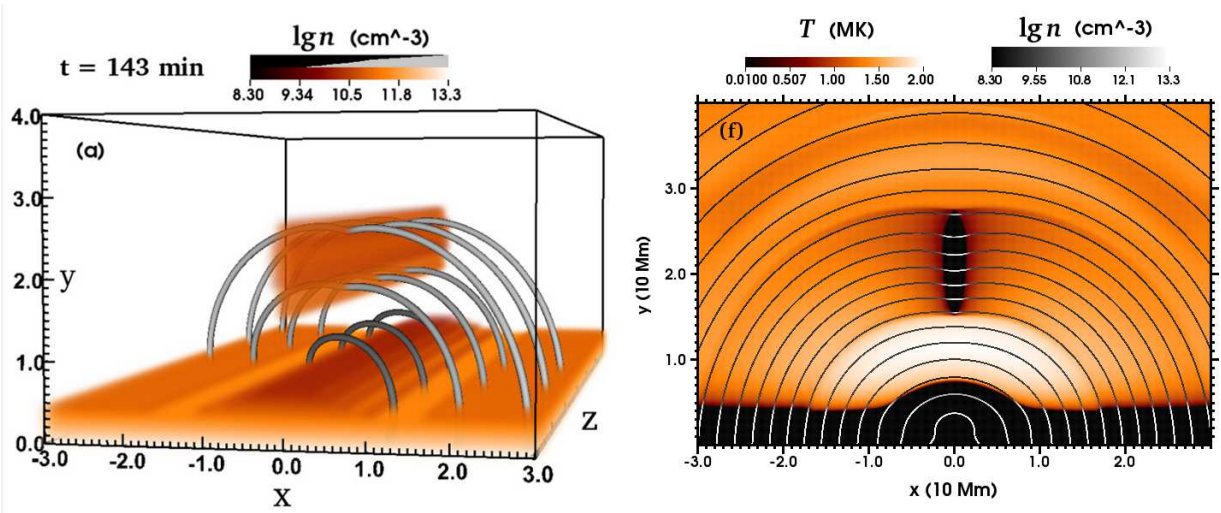}
\caption{A K--S prominence at $t=$ 143 mins. (a) 3D impression of
the prominence shown by the number density and selected magnetic 
field lines. The bar above the color bar shows the opacity of 
corresponding colors. (b) the temperature and the projected field 
lines colored by the local density.
}
\label{fig:fin}
\end{figure}

As the prominence grows fatter and heavier, the field lines 
penetrating the prominence are gradually bend downwards forming 
vertically aligned dips. This is shown in Figure \ref{fig:fin}(a),  
a 3-dimensional (3D) illustration of the vertical `sheet-like' 
prominence and selected field lines at $t=143$ minutes. The right 
panel of Figure \ref{fig:fin} gives the corresponding 2D projected 
image. At this moment, the prominence has a vertical extension of 12 
Mm from its bottom at 15.3 Mm to its top at 27.3 Mm. The horizontal 
width of the prominence is about 2.7 Mm at the middle and very thin 
at top and bottom edges. In the prominence, the temperature is about 
18600 K, the density varies from $7.3\times10^{10}$ at the bottom to 
$4.3\times10^{10}$ cm$^{-3}$ at the top, and the plasma beta changes
from 0.22 to 0.47 as the magnetic field strength changes from 6.8 G at
the bottom to 3.7 G at the top. The coronal loops below the prominence 
are heated to nearly 2 MK, which is hotter than the other coronal area 
($\sim$1.4 MK). The transition region above the neutral line has risen 
to a height of about 7 Mm, which is similar to previous works 
\citep{Lion01,Mok05}.

\begin{figure}
\includegraphics[width=6.5in]{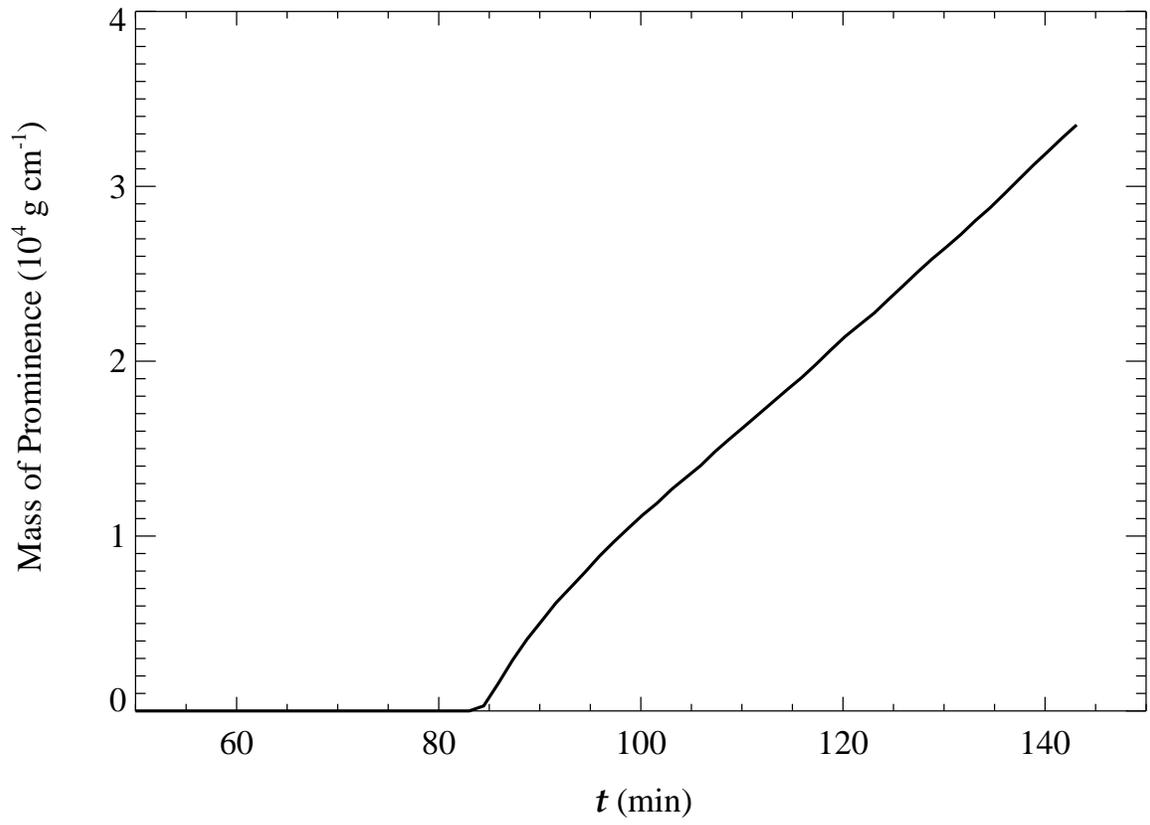}
\caption{Temporal evolution of the prominence mass.}
\label{fig:mass}
\end{figure}

The total mass of the prominence is increasing as the localized 
chromospheric heating is kept. In order to quantify the prominence 
mass, we integrate the plasma  with density larger than $4.2\times
10^{10}$ cm$^{-3}$ within a square box which surrounds the prominence. 
Within the first 10 minutes, the prominence mass grows nonlinearly 
with a mean rate of about 764 g cm$^{-1}$ min$^{-1}$. Later the 
prominence mass grows linearly, with a rate of about 519 g cm$^{-1}$ 
min$^{-1}$. The temporal evolution of the total mass of the 
prominence in a unit length in the $z$-direction is quantified in 
Figure \ref{fig:mass}. 

\begin{figure}
\includegraphics[width=6.5in]{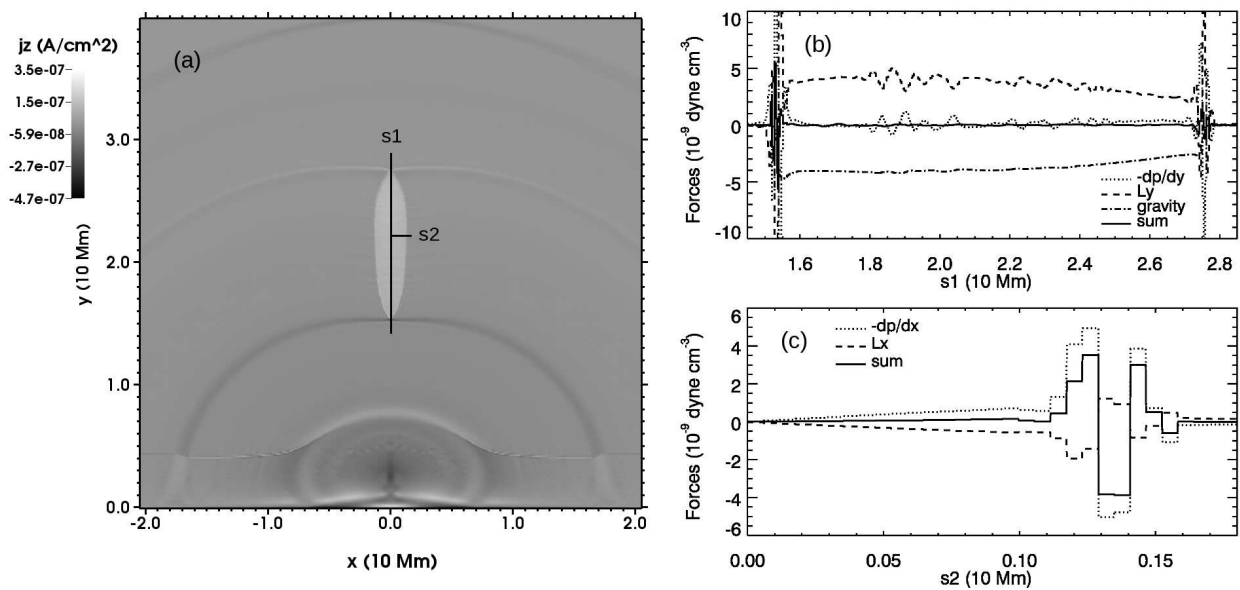}
\caption{Panel (a): Distribution of $z$-component of the current density 
($j_z$); Panel (b): Distributions of various forces along the vertical
slice s1 in panel (a), including the gas pressure gradient ({\it dotted
line}), Lorentz force ({\it dashed line}), gravity ({\it dashed dotted
line}), and their sum ({\it solid line}); Panel (c): Distributions of
various forces along the horizontal slice s2 in panel (a), including 
pressure gradient ({\it dotted line}), Lorentz force
({\it dashed line}), and their sum ({\it solid line}).
}
\label{fig:force}
\end{figure}

The initial magnetic field was force-free, with the current being 
parallel to the magnetic field. After the formation of the prominence, 
the current is locally increased in the prominence region. The 
$z$-component of the current density in the prominence is positive and 
is significantly stronger than in the surrounding corona, as shown
by Figure \ref{fig:force}(a). The magnetic field in the plane is
pointing from the left to the right, so the Lorentz force in the
prominence is pointing upward and is able to balance the gravity of 
local dense plasma. This is convincingly demonstrated by 
comparing the gravity, Lorentz force, gas pressure gradient, and their
sum, through the prominence structure. We check this first along a 
central vertical line s1 through the prominence (see Figure 
\ref{fig:force}(a)). The distributions of these forces are displayed 
in Figure \ref{fig:force}(b). Near the prominence-corona transition 
regions (PCTR) at 15.3 Mm and 27.6 Mm heights, the gas pressure 
gradient and Lorentz force fluctuate rapidly, as this transition 
region is complicated by the local thermodynamics driven by the 
radiative losses combined with field-aligned heat conduction. In this 
thin PCTR layer, forces are not in balance exactly, but the average 
value of the resultant force to remain close to zero. However, in the
prominence body, the dominant Lorentz force and gravity nearly 
balance perfectly, except for small fluctuations of the 
Lorentz force, which are then compensated by opposite fluctuations
of the pressure gradient. The resultant force vanishes throughout the
prominence body, as indicated by the solid line, realizing 
force-balance in the vertical direction. In the horizontal direction, 
the distributions of the Lorentz force, the pressure gradient, and 
their resultant force along the slice s2 are plotted in panel (c) of 
Figure \ref{fig:force}. We find that inside the prominence ($0<x<1.1$ 
Mm), the force of the pressure gradient points outwards and almost 
balanced by the Lorentz force, which is pointing to the center of the 
prominence. Hence the horizontal balance is realized by a magnetic 
pinching. The strength of these forces increases from the center to 
the edge of the prominence, and a small resultant force points 
outwards. In the thin PCTR region, forces fluctuate again indicating 
the detailed thermodynamic processes at play there.
In the end, the mass of the prominence would be saturated, after which 
any newly-formed condensation would fall aside from the corresponding 
magnetic dips, forming the drainage of cool material.

\section{CONCLUSION}\label{conclusion}
In this Letter, we simulate the formation of a normal-polarity 
quiescent prominence in a magnetized coronal arcade by 
chromospheric heating for the first time in a realistic 
multi-dimensional magnetic configuration. There should be no 
fundamental difference in the inverse-polarity configuration, 
which requires 3D simulations, as far as the radiatively driven 
condensation is concerned. The magnetic dips supporting the 
prominence mass against gravity are self-consistently formed in 
an overall low-beta environment. 
Therefore, the magnetic dips in our model are a consequence of 
the prominence formation. Our simulation captures many details 
relevant to quiescent prominence models, and closely resembles many
observational features. The prominence body, elongated along the 
invariant direction, is situated above the magnetic neutral line 
and the magnetic field through the prominence makes a finite 
angle to its axis. Our model naturally produces a vertical `sheet-like' 
prominence rather than multiple threads aligned along the magnetic 
field lines. As long as the chromospheric heating is active, 
the prominence grows in horizontal and vertical size, while the 
mass-loaded arcade field loops realize a force balance between 
Lorentz force and gravity throughout the prominence body. Our 
simulation radically improves all earlier evaporation-condensation 
studies, which assume rigid one-dimensional field line shapes 
along which catastrophic cooling scenarios have been studied in 
parametric detail, invariably leading to field-aligned prominence 
threads. Similar parametric studies will be required in our fully 
multi-dimensional settings, to determine their impact on the 
macroscopic parameters like prominence width, height and total mass. 

This study can guide future observations dedicated to find direct 
evidence of plasma condensation and mass-supply mechanism in 
prominences and ultimately uncover these mysteries of prominences. 
Our model can act as a starting point for future studies of 
Rayleigh--Taylor instability development in quiescent prominences, 
extending recent local box studies \citep{Hill11}, by allowing for 
true sheared field configurations. We intend to contrast synthetic 
views on the obtained prominence structures with modern observations, 
and will investigate how spatio-temporally varying heating conditions 
may give rise to multiple condensation sites and filament threads. 

\acknowledgments
C.X. thanks R.L. Jiang, F. Chen, and Y. Guo for discussions. This 
research was supported by NSFC under grants 11025314, 10878002,
and 10933003, and NKBRSF under grant 2011CB811402. Computations used 
the IBM Blade Center HS22 Cluster at the HPC Center of Nanjing
University of China. R.K. acknowledges funding from EC Seventh 
Framework Programme (FP7/2007-2013) under grant agreement SWIFF
(project nr 263340, www.swiff.eu).

\clearpage

\clearpage

\clearpage

\clearpage

\clearpage

\end{document}